\documentclass[prc,showpacs,showkeys,aps,epsfig]{revtex4}

\usepackage{psfig}

\def\lsim{\raise0.3ex\hbox{$<$\kern-0.75em\raise-1.1ex\hbox{$\sim$}}}

\def\gsim{\raise0.3ex\hbox{$>$\kern-0.75em\raise-1.1ex\hbox{$\sim$}}}

\newcommand{\be}{\begin{equation}}

\newcommand{\ee}{\end{equation}}

\def\alphaem{\alpha_{em}}

\def\beq{\begin{equation}}

\def\eeq{\end{equation}}

\def\beqa{\begin{eqnarray}}

\def\eeqa{\end{eqnarray}}

\newcommand{\rr}{\mbox{\boldmath $r$}}

\newcommand{\rb}{\mbox{\boldmath $b$}}

\def\gappeq{\mathrel{\rlap {\raise.5ex\hbox{$>$}}

{\lower.5ex\hbox{$\sim$}}}}

\def\lappeq{\mathrel{\rlap{\raise.5ex\hbox{$<$}}

{\lower.5ex\hbox{$\sim$}}}}

\def\Toprel#1\over#2{\mathrel{\mathop{#2}\limits^{#1}}}

\begin{document}

\title{Could saturation effects  be visible in  a future electron-ion collider?}
\author{ E.R. Cazaroto$^1$, F. Carvalho$^1$,  V.P. Gon\c{c}alves$^2$, and  F.S. Navarra$^1$}
\affiliation{$^1$Instituto de F\'{\i}sica, Universidade de S\~{a}o Paulo,
C.P. 66318,  05315-970 S\~{a}o Paulo, SP, Brazil\\
$^2$High and Medium Energy Group (GAME), \\
Instituto de F\'{\i}sica e Matem\'atica,  Universidade
Federal de Pelotas\\
Caixa Postal 354, CEP 96010-900, Pelotas, RS, Brazil\\}

\begin{abstract}
We expect to observe parton saturation in  a future electron - ion collider.
In this letter we discuss this expectation in  more detail considering two different 
models which are in good agreement with the existing experimental data on  nuclear structure 
functions. In particular,  
we study the predictions of saturation effects in  electron - ion collisions at high energies, 
using a generalization for nuclear targets of the b-CGC model, which describes the $ep$ HERA  
quite well. We estimate the total, longitudinal and charm  structure functions in the dipole 
picture and compare them with the predictions obtained using collinear factorization and modern 
sets of  nuclear parton distributions. Our results show that  inclusive observables are 
not very useful in the search for saturation effects. In the small $x$ region they are very 
difficult to disentangle from the predictions of the  collinear approaches . This happens 
mainly because of the large uncertainties in the latter. On the other hand, our results 
indicate that the contribution of diffractive  processes to  the total cross section  is 
about 20 \% at large $A$ and small $Q^2$, allowing for a detailed study of diffractive 
observables. 
The study of diffractive processes becomes essential to observe parton  saturation.

\end{abstract}



\maketitle
\vspace{1cm}


The search for signatures of parton saturation effects has been  subject of an active 
research in the last years (for recent reviews see, e.g. \cite{iancu_raju}). It has been observed that HERA data in the  small $x$ and low $Q^2$ region  can be successfully described with saturation models 
\cite{GBW,bgbk,kowtea,iim,fs}. Moreover, the experimentally measured total  cross 
sections   present the property of geometric scaling \cite{scaling} and the observed  supression of high $p_T$ hadron yields at forward rapidities in $dAu$ collisions at RHIC  \cite{brahms} follows the behavior predicted in the framework of the color glass condensate (CGC) formalism  \cite{jamal,kkt,dhj,gkmn,buw}. All these results provide some 
evidence for saturation at HERA and RHIC. However, more definite conclusions are not possible 
due to the small value of the saturation scale in the kinematical range of HERA and due to 
the complexity of $dAu$ collisions, where we need to consider the substructure of the projectile 
and the target, as well as the fragmentation of the produced partons. So far, other models 
(without saturation included) are able to describe the same set of data (see e.g. Refs. 
\cite{forshaw_twopomeron,hwa}). In order to discriminate between these different models and 
test the CGC physics, it would be very important to consider an alternative search. To this 
purpose, the future electron-nucleus colliders are ideal 
\cite{raju_ea1,raju_ea2,raju_ea4}, because they can probably 
determine whether parton distributions saturate or not. 

After the eRHIC was proposed \cite{raju_ea2}, it became crucial to have some quantitative 
estimates of the impact of saturation effects on observables. Some of these estimates can be 
found in \cite{kgn1,kgn2,niko,raju_ea4}. In particular, in  \cite{kgn1} we calculated  several inclusive observables  ($F^A_2$, $F^A_L$, $F^{c,A}_{2}$ and their logarithmic slopes), while in 
\cite{kgn2} the  behavior of the 
diffractive structure function $F_2^{D(3)}$ and the  diffractive cross section $\sigma^{diff}$ were studied in detail.   An interesting conclusion from Ref. \cite{kgn2} was related to the growth of $R_{\sigma}= {\sigma^{diff}} / {\sigma^{tot}}$ with the atomic number of the target, especially in the small $x$ and low 
$Q^2$ region. This ratio 
could be as large as $0.3 - 0.4$.  This is very large compared to the  corresponding ratio in 
$ep$ collisions 
which is of the order of $0.10 - 0.15$.  Such a large value of $R_{\sigma}$ was anticipated long 
ago in Ref.  \cite{niko95} but  the investigation of this enhancement was not 
further developed. Literally, our result implies that in 30 to 40 \% of the $eA$ collisions 
the nucleus escapes intact! This phenomenon is so spectacular that it deserves further 
investigation. 

Our goal in this letter is twofold. Firstly, to improve our previous estimates for the inclusive observables at $eA$ colliders considering a more realistic model for the nuclear dipole cross section which allows us to describe the scarce experimental data on nuclear structure functions. 
This can give us more confidence about the behavior predicted by the saturation model for $F_2^A$, $F_L^A$ and $F_2^{c,A}$ at the small $x$. Secondly, to compare these predictions with those obtained using collinear factorization and the parametrizations of the nuclear parton distributions. 
These improved calculations should  allow us to give a partial answer to the question posed in 
the title of the paper. 

In order to study the behavior of the observables of  deep inelastic scattering (DIS) at small 
$x$ and to include  saturation effects, it is useful to describe the  photon-hadron scattering in the dipole frame, in which most of the energy is carried by the hadron, while the  photon  has
just enough energy to dissociate into a quark-antiquark pair before the scattering. This description contrasts  with the usual description in the infinite momentum frame  of the hadron, based on collinear factorization, where the photon scatters a sea quark, which is typically emitted  by the small-$x$ gluons in the hadron. 
The QCD description of DIS at small $x$ can be
interpreted as a two-step process \cite{nik}. The virtual photon (emitted
by the incident electron) splits into a $q \bar{q}$ dipole,   with transverse separation 
$\rr$, which subsequently interacts with the target.  In terms
of  virtual photon-target cross sections  $\sigma_{T,L}$
for the transversely and longitudinally  polarized photons, the nuclear  $F_2$ structure function is 
given by $F_2^A(x,Q^2)\,=\,\frac{Q^2}{4 \pi^2 \alphaem} \,\sigma_{tot}$, with  \cite{nik}
\be
\label{eq:1}
\sigma_{tot} = \sigma_T\,+\,\sigma_L \,\,\,\mbox{and}\,\,\,\sigma_{T,L}\,=\,  \int d^2{\rr}\, dz\, |\Psi_{T,L}(\rr,z,Q^2)|^2\,\, \sigma_{dip}^A(x,\rr),
\ee
where $\Psi_{T,L}$ is the light-cone  wave function of the virtual photon
and $\sigma_{dip}^A$ is the   dipole nucleus cross section
describing  the interaction of the $q\bar{q}$  dipole with the nucleus target.  In
equation (\ref{eq:1})
$\rr$ is the transverse separation of the $q\bar{q}$ pair
and $z$ is the photon  momentum fraction carried by the quark (for details see  e.g. Ref. \cite{PREDAZZI}). Moreover, the nuclear longitudinal structure function is given by 
$F_L^A(x,Q^2)\,=\,(Q^2/4 \pi^2 \alphaem) \,\sigma_L$ and the charm component of the nuclear 
structure function $F_2^{c,A}(x,Q^2)$ is obtained directly from Eq. (\ref{eq:1}) isolating the charm flavor and considering $m_c = 1.5$ GeV.

The main input for the calculations of  inclusive observables in the dipole picture is $\sigma_{dip}^A(x,\rr)$ which is determined by the QCD dynamics at small $x$. In the eikonal approximation it is  given by:
\begin{equation} 
\sigma_{dip}^A(x, \rr) = 2 \int d^2 \rb \,  {\cal N}^A(x, \rr, \rb)
\label{sdip}
\end{equation}
where $ {\cal N}^A(x, \rr, \rb)$ is the forward scattering amplitude for a dipole with size 
$\rr$ and impact parameter $\rb$ which encodes all the
information about the hadronic scattering, and thus about the
non-linear and quantum effects in the hadron wave function. It  can be obtained by solving the BK (JIMWLK) evolution equation in the rapidity $Y \equiv \ln (1/x)$  \cite{iancu_raju}. 
Many groups have studied the numerical solution of the BK equation \cite{solBK}, but several improvements are still necessary  before using the solution in the calculation of  observables. In particular, one needs to include the next-to-leading order corrections into the evolution equation and perform a global analysis of all small $x$ data. It is a program in progress (for recent results see \cite{alba}). In the meantime it is necessary to use phenomenological models for 
$ {\cal N}^A$ which capture the most essential properties of the solution.
 
 In \cite{kgn1,kgn2} we assumed that the impact parameter 
dependence of $ {\cal N}^A$ can be factorized as ${\cal N}^A(x, \rr, \rb) = {\cal N}^A(x, \rr) S(\rb)$ and proposed a generalization for the nuclear case of the IIM model \cite{iim} which was,  
at that time,   the most 
 sophisticated one. In other words, we have disregarded the impact parameter dependence of the scattering amplitude and assumed that the dipole nucleus scattering amplitude is related to 
the dipole proton one  by a simple modification in the saturation scale: $Q_{s,p}^2(x) \rightarrow Q_{s,A}^2 =  A^{\frac{1}{3}} \times  Q_{s,p}^2(x)$. This naive model is useful to obtain some idea about the magnitude of the saturation effects in $eA$ processes in comparison to the linear case. However, in order to get more reliable predictions we should use a phenomenological model which describes the current scarce experimental data on the nuclear structure function as well as includes the  impact parameter dependence in the dipole nucleus cross section. A  model which satisfies these 
requirements was proposed some years ago in Ref. \cite{armesto}. In this model the forward dipole-nucleus amplitude was   parametrized  as follows
\begin{eqnarray}
{\cal{N}}^A(x,\rr,\rb) = 1 - \exp \left[-\frac{1}{2}A \,T_A(\rb) \, \sigma_{dip}^p(x,\rr^2)\right] \,\,,
\label{enenuc}
\end{eqnarray}
where $T_A(\rb)$ is the nuclear profile function, which is 
obtained from a 3-parameter Fermi distribution for the nuclear
density normalized to unity (for details see, e.g., Ref. \cite{vicmag_hq}).
The above equation, based on the Glauber-Gribov formalism \cite{gribov},  sums up all the multiple elastic rescattering diagrams of the $q \overline{q}$ pair
and is justified for large coherence length, where the transverse separation $r$ of partons in the multiparton Fock state of the photon becomes a conserved quantity, {\it i. e.} the size of the pair $r$ becomes eigenvalue
of the scattering matrix. It is important to emphasize that for very small values of $x$, other diagrams beyond the multiple Pomeron exchange considered here should contribute ({\it e.g.} Pomeron loops) and a more general approach for the high density (saturation) regime must be considered. However, we believe that this approach allows us to estimate the magnitude of the high density effects in the RHIC and LHC kinematic range.

In  \cite{armesto} the author has assumed that  $\sigma_{dip}^p$ was given by  the GBW model.
However, in the last years an intense activity in the area resulted  in  more sophisticated 
dipole proton cross sections, which had more  theoretical 
constraints and which were able to give a better description of the more recent HERA data 
\cite{kkt,dhj,kmw,gkmn,watt,buw}. In what follows we will use the b-CGC model proposed in Ref. \cite{kmw}, which improves the IIM model 
 \cite{iim} with  the inclusion of   the impact parameter dependence in the dipole proton cross sections. The parameters of this model were recently fitted to describe the current HERA data \cite{watt}.  Following \cite{kmw} we have:
\begin{equation}
\sigma_{dip}^p (x,\rr^2) \equiv \int \, d^2 \bar{\rb} \, \frac{d \sigma_{dip}^p}{d^2  \bar{\rb}} 
\label{new_iim}
\end{equation}
where 
\begin{eqnarray}
\frac{d \sigma_{dip}^p}{d^2 \bar{\rb}} = 2\,\mathcal{N}^p(x,\rr,\bar{\rb}) =  2 \times 
\left\{ \begin{array}{ll} 
{\mathcal N}_0\, \left(\frac{ r \, Q_{s,p}}{2}\right)^{2\left(\gamma_s + 
\frac{\ln (2/r Q_{s,p})}{\kappa \,\lambda \,Y}\right)}  & \mbox{$r Q_{s,p} \le 2$} \\
 1 - \exp^{-a\,\ln^2\,(b \, r \, Q_{s,p})}   & \mbox{$r Q_{s,p}  > 2$} 
\end{array} \right.
\label{eq:bcgc}
\end{eqnarray}
with  $Y=\ln(1/x)$ and $\kappa = \chi''(\gamma_s)/\chi'(\gamma_s)$, where $\chi$ is the 
LO BFKL characteristic function.  The coefficients $a$ and $b$  
are determined uniquely from the condition that $\mathcal{N}^p(x,\rr)$, and its derivative 
with respect to $rQ_s$, are continuous at $rQ_s=2$. 
In this model, the proton saturation scale $Q_{s,p}$ now depends on the impact parameter:
\begin{equation} 
  Q_{s,p}\equiv Q_{s,p}(x,\bar{\rb})=\left(\frac{x_0}{x}\right)^{\frac{\lambda}{2}}\;
\left[\exp\left(-\frac{\bar{b}^2}{2B_{\rm CGC}}\right)\right]^{\frac{1}{2\gamma_s}}.
\label{newqs}
\end{equation}
The parameter $B_{\rm CGC}$  was  adjusted to give a good 
description of the $t$-dependence of exclusive $J/\psi$ photoproduction.  
Moreover the factors $\mathcal{N}_0$ and  $\gamma_s$  were  taken  to be free. In this 
way a very good description of  $F_2$ data was obtained. 
The parameter set  which is going to be used here is the one presented in the second line of
Table II of \cite{watt}:  $\gamma_s = 0.46$, $B_{CGC} = 7.5$ GeV$^{-2}$,
$\mathcal{N}_0 = 0.558$, $x_0 = 1.84 \times 10^{-6}$ and $\lambda = 0.119$.

\begin{figure}
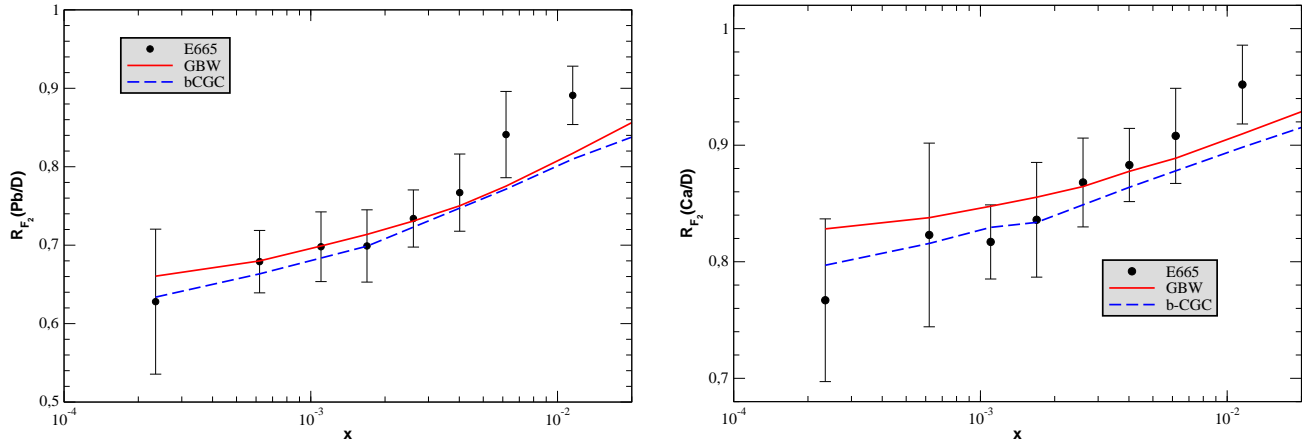

\vspace*{0.5cm}
\centerline{
\begin{tabular}{ccc}
{\psfig{figure=ratio_pbD2.eps,width=8.3cm}} & \,\,\,\,\,\, & {\psfig{figure=ratioCaD2.eps,width=8.3cm}} 
\end{tabular}}
\caption{Ratio $R_{F_2} \equiv 2 F_2^A / A F_2^D $ for $A = Pb$ and $A = Ca$. Although joined by a line the results of the saturation models have been computed for the $(\langle x \rangle, \, \langle Q^2 \rangle)$ of the data points  from the E665 Collaboration \cite{e665}. }
\label{fig1}
\end{figure}

In Fig. \ref{fig1} we compare with the E665 data \cite{e665} the predictions for the ratio $R_{F_2} = 2 F_2^A / A F_2^D $ obtained using the b-CGC model as input in our calculations. For comparison we also show the predictions obtained using the GBW model. As expected, both models fail 
to describe the large $x$ region. However, both models describe quite well the scarce experimental data in the region of small values of $x$ and low $Q^2$. The basic difference between the predictions occur in the normalization of the ratio, with b-CGC model predicting a large nuclear effect at small $x$.  It is important to emphasize that after the choice of the model for the dipole proton cross section our predictions are parameter free.  Since our model for the dipole nuclear dipole cross section, Eq. (\ref{enenuc}),  describes reasonably well the experimental data for the nuclear structure
function, we feel confident to compare its predictions with those obtained using the collinear factorization as well as in extending it to the calculation of the longitudinal and charm structure functions.

\begin{figure}
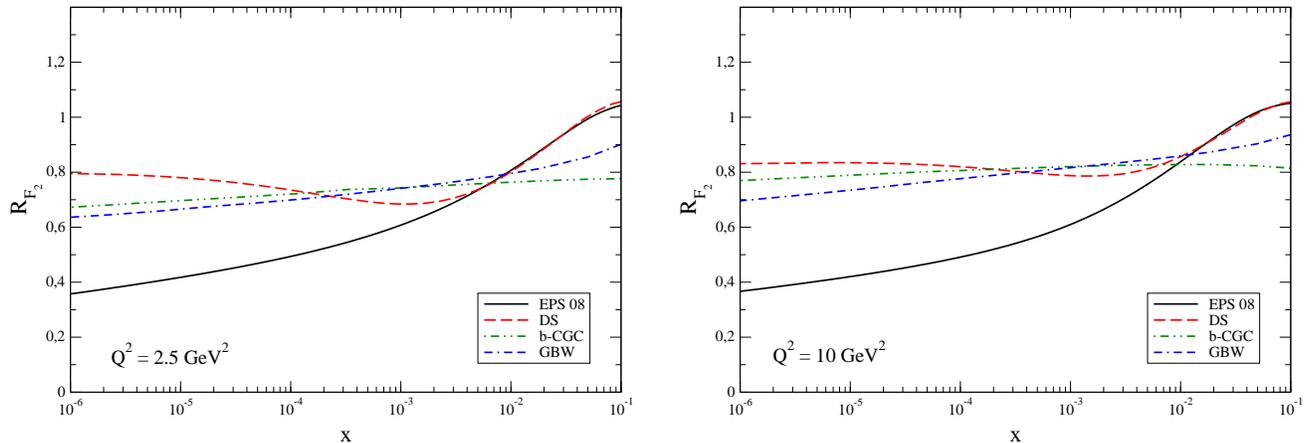

\vspace*{0.5cm}
\centerline{
\begin{tabular}{ccc}
{\psfig{figure=Rf2_eps_ds_bcgc_gbw_2_5.eps,width=8.3cm}} & \,\,\,\,\,\, & {\psfig{figure=Rf2_eps_ds_bcgc_gbw_10.eps,width=8.3cm}} 
\end{tabular}}
\caption{Comparison between the predictions for $R_{F_2}$ of the saturation models (b-CGC and GBW) and the collinear ones (DS and EPS08) for two different values of $Q^2$ and $A = Pb$.}
\label{fig2}
\end{figure}

In Fig. \ref{fig2} we present  our predictions for the $x$ dependence of the ratio $R_{F_2}$ for two different values of the photon virtuality $Q^2$. The two saturation models predict a similar 
behavior in the small $x$ region, yielding  values of $\approx$ 0.7 for $x = 10^{-4}$ and $Q^2 = 2.5$ GeV$^2$. This value grows to $\approx$ 0.8 for  $Q^2 = 10$ GeV$^2$  For comparison we also show the predictions obtained using the collinear factorization and nuclear parton distributions resulting from the global analysis of the nuclear experimental data using the DGLAP evolution equation  (for a recent discussion see \cite{ccgn}). Here we consider two different sets of nuclear parton distributions: the DS \cite{ds04} and EPS 08 \cite{eps08} nuclear 
parametrizations. They represent a lower and an upper bound for the magnitude of the nuclear effects, respectively. In particular, in the analysis of Ref. \cite{eps08} the authors  include data on hadron 
production in the forward region at  RHIC \cite{brahms}. As a result the amount of nuclear
shadowing   
is considerably larger than the one obtained in other parametrizations. It is important to emphasize that the collinear predictions {\it do  not} include dynamical  effects associated to non-linear (saturation) physics, since they are based on the linear DGLAP dynamics. Therefore, the comparison between the saturation and collinear predictions could, in principle, tell us if the 
observable considered can be used to discriminate between linear and non-linear dynamics. 
The results shown in Fig. \ref{fig2} demonstrate that this is not the case of the nuclear structure function. Although the predictions of the two collinear models are similar for large $x$ ($\ge 10^{-2}$), where there exist experimental data, the predicted behavior at small $x$ is very 
distinct. This difference is a  consequence of  the choice of different experimental data used 
in the global analysis and of the assumptions related to the behavior of the nuclear gluon density \cite{ccgn}. The difference between the collinear predictions is so large at small $x$ that it 
is not possible to extract any information about the presence or not of new QCD dynamical 
effects from  the study of the nuclear structure function. 

\begin{figure}
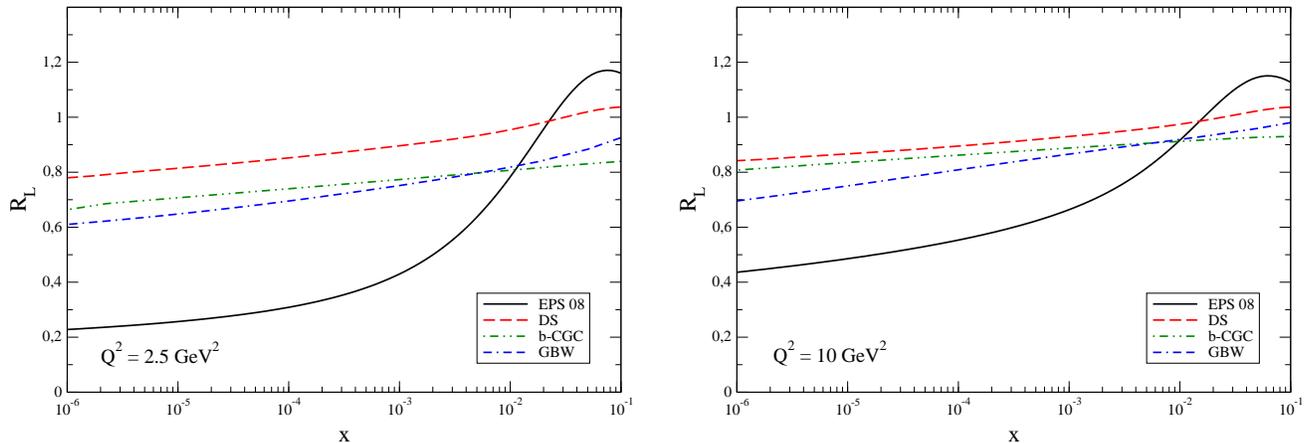

\vspace*{0.5cm}
\centerline{
\begin{tabular}{ccc}
{\psfig{figure=RL_eps_ds_bcgc_gbw_2_5.eps,width=8.3cm}} & \,\,\,\,\,\, & {\psfig{figure=RL_eps_ds_bcgc_gbw_10.eps,width=8.3cm}} 
\end{tabular}}
\caption{Comparison between the predictions for $R_{L} \equiv 2 F_L^A / A F_L^D $ of the saturation models (b-CGC and GBW) and the collinear ones (DS and EPS08) for two different values of $Q^2$ and $A = Pb$.}
\label{fig3}
\end{figure}

Let us now consider the behavior of the nuclear longitudinal structure function. 
It is believed that this observable can be used to constrain the QCD dynamics at small $x$ 
in $ep$ collisions at HERA (See, e.g. \cite{vicmag_fl}). One of our goals is to verify if this 
assumption is also valid in $eA$ collisions. Our results are show in Fig. \ref{fig3}. 
The  predictions of saturation models are, as in the case of   $R_{F_2}$, contained within the 
uncertainty range of the collinear models. 
On the other hand, the difference between the two collinear models is bigger. 
In this case the results for large $x$ are  distinct too. This is due to the 
strong  dependence of $F_L^A$ on the nuclear gluon distribution, which is very different in 
the two models considered \cite{ccgn}.

\begin{figure}
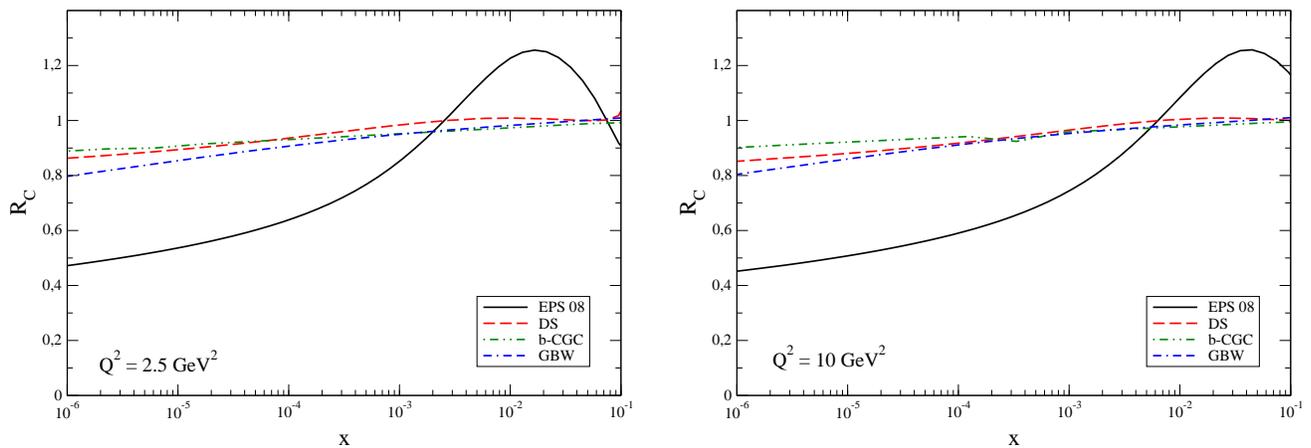

\vspace*{0.5cm}
\centerline{
\begin{tabular}{ccc}
{\psfig{figure=Rc_eps_ds_bcgc_gbw_2_5.eps,width=8.3cm}} & \,\,\,\,\,\, & {\psfig{figure=Rc_eps_ds_bcgc_gbw_10.eps,width=8.3cm}} 
\end{tabular}}
\caption{Comparison between the predictions for $R_{C} \equiv 2 F_2^{c,A} / A F_2^{c,D}$ of the saturation models (b-CGC and GBW) and the collinear one (DS and EPS08) for two different values of $Q^2$ and $A = Pb$.}
\label{fig4}
\end{figure}

Finally, we show in Fig. \ref{fig4} our results for the ratio $R_{C} \equiv 2 F_2^{c,A} / A F_2^{c,D}$, which is determined by the   charm component of the nuclear structure function. In this 
case the saturation model predictions and the DS collinear ones are similar, 
while the EPS08 one is very distinct. Similarly to the $R_L$ case, this behavior is associated 
to the large magnitude of nuclear effects present in the nuclear gluon distribution 
predicted by the EPS08 parametrization, which has the strongest shadowing.  We did not
include in the plots the somewhat older but very well known EKS parametrization. Its predictions 
for $R_{F_2}$, $R_{L}$ and $R_{C}$ would always lie between the GBW and EPS08 curves. So we can 
say that, even neglecting the EPS08 parametrization (which is the most extreme), the predictions 
of saturation models overlap with those from collinear models.

Let us summaryze our results for inclusive observables $F_2^A$, $F_L^A$ and $F_2^{c,A}$, 
which should be measured in the first run of a future electron-ion collider. The  conclusion 
that we can draw from the previous  figures is  that the two dipole models yield similar 
predictions for the structure functions and they fall inside the range of predictions of the EPS 
and DS parametrizations. The latter differ among each other because of the  large 
freedom inherent to global data analysis. 

Although these observables are  sensitive to saturation effects, as shown in \cite{kgn1} by 
the comparison between the predictions of the full (linear $+$ non-linear contributions) 
dipole models and their  corresponding  purely linear versions, it is not yet possible to draw 
any firm conclusion concerning the QCD dynamics from  inclusive quantities. 
This is a pessimistic partial answer for the question posed on the title, which implies that 
in order to discriminate the saturation effects we should consider less inclusive observables.

\begin{figure}
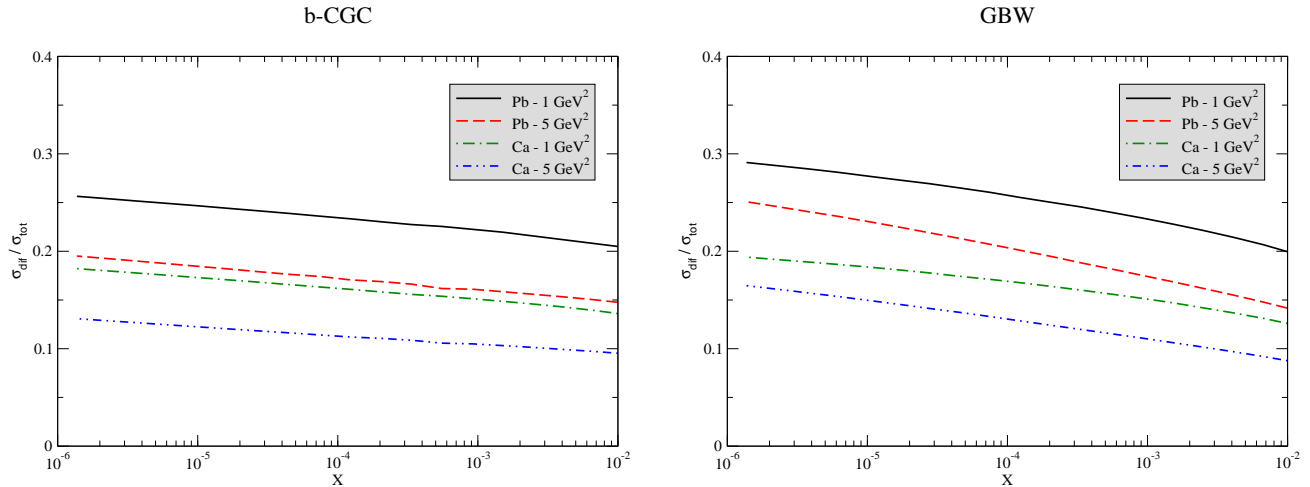

\vspace*{0.5cm}
\centerline{
\begin{tabular}{ccc}
{\psfig{figure=razao_bcgc_dif.eps,width=8.3cm}} & \,\,\,\,\,\, & {\psfig{figure=razao_gbw_dif.eps,width=8.3cm}} 
\end{tabular}}
\caption{Ratio of  diffractive to total cross sections, $R_{\sigma}$, as a function of $x$ with the b-CGC and GBW models. Long dashed and solid lines are for Pb targets at $Q^2=5.0$ and $1.0$ GeV$^2$, respectively. 
Dot-dot-dash and dot-dash lines are the same for Ca targets.}
\label{fig5}
\end{figure}

An alternative can be the study of the logarithmic slopes of the structure functions, as already discussed in \cite{kgn1,vic_slope}. As pointed out in \cite{vic_slope}, the logarithmic slope of $F_2^A$ can be useful to address the boundary between the linear and saturation  regimes, due to the $A$ dependence present in the nuclear saturation scale (See also \cite{kgn1}). Another possibility is the study of observables measured in diffractive deep inelastic scattering (DDIS), since  the total diffractive cross section is much more sensitive to large-size dipoles than the inclusive one \cite{GBW,kgn2}. Basically, the saturation effects  screen large-size dipole (soft) contributions, so that a fairly large
fraction of the cross section is hard and hence eligible for a perturbative treatment. This was the main motivation of  Ref. \cite{kgn2}, where we have computed observable quantities
like $R_{\sigma} = \sigma_{diff}/\sigma_{tot}$ and  $F_{2}^{D(3)}$ in the dipole picture using a naive saturation model. One of the main conclusions was  related to the growth of $R_{\sigma}$ with the atomic number of the target, especially in the small $x$ and low 
$Q^2$ region, where the ratio could be as large as $0.3 - 0.4$. Although we postpone for a future publication a detailed analysis of the diffractive observables using the b-GCC model discussed in this letter, it is important to verify if the diffractive contribution for the total cross section is still large when estimated using a more realistic saturation model. Following \cite{raju_ea4} we estimate the total diffractive cross section as follows
\begin{equation}
\sigma_{diff} = \sigma^D_{L} \, + \,  \sigma^D_{T} \,\,\,\mbox{and}\,\,\,\sigma^D_{L,T} = \frac{1}{4}\,\int d^2{\rr}\, dz\, |\Psi_{T,L}(\rr,z,Q^2)|^2 \int \, d^2 {\rb} \, \left( \frac{d \sigma_{dip}^A}{d^2  {\rb}} \right)^2 
\label{sigdif}
\end{equation}
where $d \sigma_{dip}^A/d^2  {\rb} = 2\,{\cal{N}}^A(x,\rr,\rb)$, with ${\cal{N}}^A$ given by Eq. (\ref{enenuc}). In Fig. \ref{fig5} we present our predictions for the ratio $R_{\sigma}$
as a function of $x$ for two values of $Q^2$ and the atomic number considering the b-CGC and GBW models.
In comparison with our previous results \cite{kgn2}, we  can observe that the inclusion of the impact parameter in the saturation model reduces  the  ratio in approximately  30 \%. However, the values predicted are still large, especially for small $Q^2$ and large $A$. Our new predictions imply that 
the ratio for $ePb$ collisions is a factor two larger than observed in $ep$ collisions at HERA. Consequently, the study of diffractive processes should be an easy task in an $eA$ collider, 
allowing for a detailed analysis of the QCD dynamics. Since the study of inclusive observables is not 
very illuminating,   the study of diffractive processes becomes fundamental in order to constrain the QCD dynamics at high energies.


As a summary, in this letter  
 we have studied the predictions of saturation physics for  electron - ion collisions at high energies, using a generalization for nuclear targets of the b-CGC model which describes the $ep$ HERA  quite well. We have  estimated the nuclear structure function $F_2^A(x,Q^2)$, as well as  the longitudinal and charm contributions and compared with the predictions obtained using collinear factorization and distinct sets of nuclear parton distributions. The basic idea is to compare the predictions from non-linear and linear QCD dynamics and verify if the  experimental analysis of these observables in the future electron - ion collider could reveal the presence of  saturation physics, as well as constrain the behavior of the saturation scale.  Our results indicate that the inclusive observables are not adequate for this purpose due to the large uncertainty present in the collinear predictions at small $x$. On the other hand, it is expected that the collinear factorization formalism fails to describe diffractive $eA$ processes, while the saturation formalism remains valid. The study of diffractive processes becomes essential to observe the saturation physics. In this letter we have estimated the contribution of these processes for the total cross section and demonstrated that it is about 20 \% at large $A$ and small $Q^2$, 
allowing for a detailed study of  diffractive observables. Our results motivate a more detailed study of the diffractive structure function as well as the leptoproduction of vector mesons in $eA$ processes. These studies are fundamental in order to get a definitive answer for the question posed on the title.
 

\begin{acknowledgments}
  This work was  partially financed by the Brazilian funding
agencies CNPq, FAPESP and FAPERGS. It is a pleasure to thank M.S. Kugeratski for 
fruitful discussions.  
\end{acknowledgments}



\end{document}